\documentclass{epl}
\usepackage{graphicx}
\usepackage[english]{babel}
\title{Phase transition in the assignment problem for random matrices}
\author{J. G. Esteve \inst{1,2} \and F. Falceto 
\inst{1,2} }
\institute{
\inst{1}  Departamento de F\'{\i}sica Te\'orica, Universidad de
 Zaragoza. Zaragoza, Spain.\\
\inst{2} Instituto de Biocomputaci\'on y F\'{\i}sica de Sistemas Complejos,
 Universidad de
 Zaragoza. Zaragoza, Spain.}
\begin{document}
\maketitle
%
%
%
%\date{}
%
%
%
\begin{abstract}
We report an analytic and
 numerical study of a phase transition in a P problem
 (the assignment problem)
 that separates two phases whose representatives are
 the simple matching problem
(an easy P problem) and the traveling salesman problem
 (a NP-complete problem). Like other phase transitions
 found in combinatoric problems
 (K-satisfiability, number partitioning) this can help to understand the
 nature of the difficulties in
 solving NP problems an to find more accurate algorithms for them.
\end{abstract}
In the theory of computational complexity,
two paradigmatic problems representative of the classes $NP$-complete
 and $P$ are the traveling salesman  and the
 assignment problem. Traveling salesman problem (TSP) 
 requires $N$ points or cities,
 with distances $d_{i,j}$ between them, for which we must find the
 closed tour of minimum length that visits each city once.
 In the assignment or bipartite matching problem (AP), we have $N$
 objects of two classes (say A and B) with distances $d_{i,j}$
 between each object $i$ of the class A and the object $j$ of the class B.
 We must assign at each object $i$ of the class A one and only one 
 object $\sigma(i)$ of the class B in such a way that the total distance 
$\sum_{i=1}^N d_{i,\sigma(i)}$ is a minimum, so we can write D(AP) as:
\begin{equation}
D(AP)={\rm Min}_{(\sigma \in S_N)} \left[ \sum_{i=1}^{N}d_{i,\sigma(i)}
\right],
\end{equation}
being $S_N$ the symmetric group. This problem can
 be seen in another way when the sets A and B are the same
 (for example cities), in this case D(AP) is the path of minimum length
 that visits each city once and is composed of so  many closed sub tours
 as needed. In other words, when $A=B$ the bipartite matching can be
 seen as a problem of $n$-traveling salesmen where each salesman must
 start and end his tour in the same city and we can use as many
 salesmen as needed in order to minimize the total length of
 all salesmen tours.
Although the configuration space for the AP problem has $N!$
 elements while that  of the TSP has {\it only} $(N-1)!$, the AP is in the
 class $P$ and can be solved in times that
 grows like $N^3$ with the number $N$ of cities.

There is another interesting limit for the AP, when we
 introduce the constrain that the number of tours must be $N/2$
 (for $N$ even) or equivalently, that the $N$ cities must be pair-wise
 matched. This limit is called simple matching problem (SMP) and in this case 
 the configuration space has $(N-1)!!$ elements and the problem remains in 
 the class $P$.

 Depending on the characteristics of the distance matrix the AP could, 
 in principle, interpolate between those situations which are near of the 
 SM problem (that is, situations where the optimal solution is composed
 approximately of $N/2$ cycles) and those which are near of the TSP problem
 (when the optimal solution is composed only of a few cycles). We shall see
 that the control parameter that governs the transition between both limits is
 the correlation between the distances $d_{i,j}$ and $d_{j,i}$ \cite{dist},
 in such a way
 that for positive correlations the optimal solution for the AP  is in
 the ``SMP regime'', whereas for anti correlated distances the solution is
 in the ``TSP regime'' \cite{Monasson}.

The aim of this work is to study the structure (with reference to
 the number of cycles) of the optimal solution for the bipartite
 matching with random distance matrices e.g. matrices whose elements
 $d_{i,j}$ are random numbers with a distribution of probability
 $\rho(d_{i,j})$. The problem of bipartite 
 matching on random matrices has been studied for many years, 
 but those studies focused on the length of the minimal 
 path ${D(AP)}$.
 For example for random matrices whose probability distribution is 
 $\rho(d_{i,j})= \exp{(-d_{i,j})}$ it was conjectured first by G. Parisi
 \cite{P1} and then proved rigorously \cite{prueb1}-\cite{prueb3} that
the average length is ${D(AP)}= \sum_{m=1}^{N} {1\over m^2}$
 with  $N$ the number of points that must be matched. For more
  general distributions the first terms of the expansion in $1/N$ of
$D(AP)$ are known 
(\cite{P21}-\cite{P4}).
 As for the TSP on random matrices with $\rho(0)=1$,
 the averaged length of the minimal tour is
 $ \overline{D(TSP)}= 2.041... + {\cal O}(1/N)$
 and the $1/N$ corrections can be computed in principle 
although {``the computations become rather long''} \cite{P5}.

\begin{figure}[h!]
%\begin{ruledtabular} \hsize=8.5cm
\begin{center}
\includegraphics[width=8.9cm,height=6.cm]{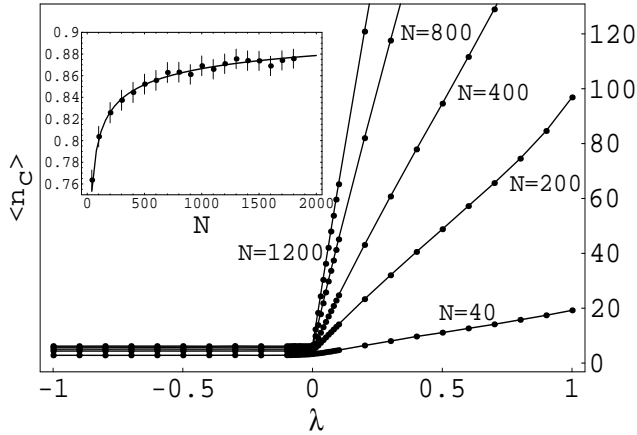}
\end{center}
\caption{\small Mean value of the number of cycles of the optimal solution for
  the assignment problem against $\lambda$ for different values
 of the number of cities  $N$. The inset is the theoretical
 (continuous line) and experimental (dots)
 mean value of the number of cycles divided by $\log (N)$ as a function of
  $N$, for $\lambda=-1$. Error bars represent three standard errors of the mean.}
\label{fig1}
%\end{ruledtabular}
\end{figure}

We shall work with random distance matrices given by
\begin{eqnarray}
\label{ranmat}
  d_{i,j}  & = &R_{ij}+\lambda R_{ji}\qquad i\not=j   \\
  d_{i,i} & = & \infty , \nonumber
\end{eqnarray}
where  $R_{ij},\ i\not=j$ are independent random variables
with uniform distribution in the interval $[0,1]$
and $\lambda \in [-1,1]$. $ d_{i,i}= \infty$ is imposed 
in order to make every salesman to travel out of his home city 
i. e. we do not allow $1$-cycles \cite{neg}.
 $\lambda =0$ corresponds to a random matrix  
 with all entries uncorrelated, 
 $\lambda =1$ is the symmetric case 
$d_{i,j}=d_{j,i}=R_{ij}+R_{ji}$ and 
 $\lambda=-1$ the antisymmetric one 
 $d_{i,j}=R_{ij}-R_{ji}=-d_{j,i}$. 
 For each value of $\lambda$ 
 we generate 
  $M$ different distance matrices (typically $M$ varies between $10^4$ and
  $10^6$)
 and we  solve for each of them the assignment problem
 using the
 algorithm of   R. Jonker and A. Volgenant \cite{JV}. Then we measure
 the mean value of the number of cycles $\langle n_{c}\rangle$, which is
 plotted in Fig. \ref{fig1}.   
%%%%%%%%%%%%%%%%%%%%%%%%%%%%%%%%%%%%%%%%%%%%%%%%%
%%%%%%%%%%%%%%%%%%%%%%%%%%
 
There we can see two very different regimes
 for the behavior of $\langle n_{c}\rangle$ as a function of $\lambda$.
When the  correlations between  $d_{i,j}$ and $d_{j,i}$ are negative  
($-1 \leq  \lambda < 0$),
$\langle n_{c}\rangle$ is (almost) constant with $\lambda$, has a  logarithmic 
dependence with $N$ i. e.  the solution for the assignment problem
is  composed typically of a few cycles, close to the TSP problem. 
This situation contrasts  with that of positively correlated
$d_{i,j}$ and $d_{j,i}$ 
($0 < \lambda \leq  1$), where $\langle n_{c}\rangle$ varies with a nearly 
linear dependence  
%(up to corrections that are more visible near $\lambda =1$)
in $\lambda$ and  $N$, 
reaching its maximum for  $\lambda=1$ with an approximate value
$\langle n_{c}\rangle_{_{\lambda=1}}\approx N/2$. At this point the
 solutions are very close to those of the SM problem in the sense that
 they are dominated by 2-cycles. 

\begin{figure}[h!]
%\begin{ruledtabular} \hsize=8.5cm
\begin{center}
\includegraphics[width=7.5cm,height=5.cm]{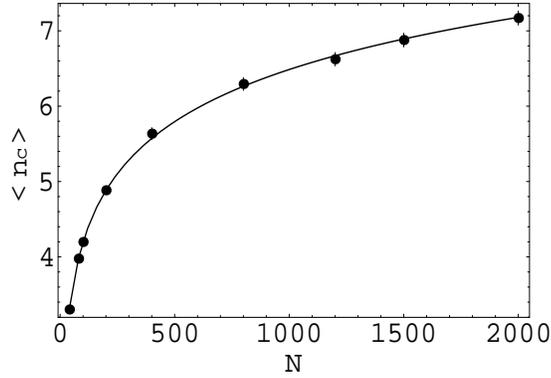} 
\end{center}
\caption{\small  Theoretical (continuous line) and experimental (dots)
 mean value of the number of cycles as a function of
  $N$, for $\lambda=0$. Again, error bars correspond to
 three standard errors of the mean.}
\label{fig2}
%\end{ruledtabular}
\end{figure}

The cases $\lambda = 0$ and $-1$ can be analyzed explicitly. 
For $\lambda=0$ the distance matrix is completely random in the sense that
off diagonal entries are equally distributed, independent random variables.
Hence all the permutations  have the same probability of being the optimal tour
 (except those which have some $1$-cycle that are excluded)
\cite{prueba} and studying the
 structure of cycles of the optimal solution is the same that to study
 the structure of cycles of the corresponding subset
 of the permutation group. 
The latter can be done with the help of 
the associated Stirling
 numbers of the first kind $d_2(N,k)$ defined as the number of
 permutations of  $N$ elements having $k$ cycles,
 all of which of length $r \geq 2 $ \cite{aSt}-\cite{Riordan}.
 Using the generating
 function of $d_2(N,k)$ we can calculate exactly the expected value of 
$n_{c}$ as:
\begin{eqnarray}
\langle n_{c}\rangle_{_{\lambda=0}} & = & -{ \left[ {d^N\over dx^N} \left( {(\log (1-x)+x )\exp (-x)
\over 1-x}\right)\right]_{x=0} \over  \left[ {d^N\over dx^N} \left( { \exp (-x)
\over 1-x}\right)\right]_{x=0}}\nonumber \\
  & = & H_N-1 + \sum_{i=1}^{\infty} C_i N^{-i},
\label{d2}
\end{eqnarray}
where $H_N=\sum_{m=1}^N(1/m)$ is the harmonic series and
the first coefficients in the expansion in (\ref{d2}) 
are $C_i= 1,$ $-1/2,$ $ -1/6,$ $
1/4, $ $8/15,$ $1/12,$ $-85/42,$ $-125/24,$ $13/90,479/10,$ $5800/33 $ for
$i=1,\dots,11$. 
In order to check the code of our simulations, in Fig. \ref{fig2}
 we plot the expected values of the mean number of cycles at $\lambda=0$  
($\langle n_{c}\rangle_{_{\lambda=0}}$)   
obtained in our simulations (dots) and compared with the
theoretical  results of (\ref{d2}) (continuous line).
 %%%%%%%%%%%%%%%%%%%%%%%%%%%%%%%%%%%%%%%%%%%%%%%%%
%%%%%%%%%%%%%%%%%%%%%%%%%% 

Note that this result is based, only, in the fact that all distances are
identically distributed independent random variables,
and is independent of the actual probability distribution.
 That is, a  change of the probability will affect, only,
 to the value of the length of the tour and not to the value of $\langle n_{c}\rangle$.
 Should we use 
random numbers also for $d_{i,i }$ (recall (\ref{ranmat}))
 then $1$-cycles would be allowed
 and the average number of cycles $\langle n_{c}\rangle^0_{_{\lambda=0}}$ 
 is exactly the Harmonic Series
 $H_N$. So for $N\to \infty$ we
 have that $\langle n_{c}\rangle_{_{\lambda=0}} = 
\langle n_{c}\rangle^0_{_{\lambda=0}} -1$,
i. e. for large values of $N$, to allow or to forbid one-cycles
 only changes in one unit the mean value of $n_{c}$.
   
%%When $\lambda$ varies between $0$ and $-0.05$,
%% with $N$ fixed, then
%% $\langle n_{c}\rangle$ decreases aproximately a constant
%% factor of $1/2$ and after that
%% (typically for $\lambda < -0.05$)  it
%% remains constant with $\lambda$.
The case $\lambda=-1$ can be studied in a similar way,
although here we do not have an exact result but an
extremely good approximation.
The key observation is that the minimal tour tends to be made of the 
smallest distances available, then 
 for anti correlated $d_{i,j}$ and $d_{j,i}$
  it is very unlikely that both enter in the minimal tour 
(because  one of them will be positive
 and the other negative adding up to zero). 
The net effect is that the anti correlations tend to
 suppress the appearance of $2$-cycles in the optimal tour and
 this effect will increase with increasing $N$.
This has been verified in the numerical simulations,
whose results are plotted in Fig. 3.  There we represent
the logarithm of the average number of 2-cycles 
in the optimal tour ($P_2$)  
as a function of $N$ for $\lambda=-1, -0.2$ and $-0.1$. 

 %%%%%%%%%%%%%%%%%%%%%%%%%%%%%%%%%%%%%%%%%%%%%%%%%
\begin{figure}[h!]
%\begin{ruledtabular} \hsize=8.5cm
\begin{center}
\includegraphics[width=7.5cm,height=5.cm]{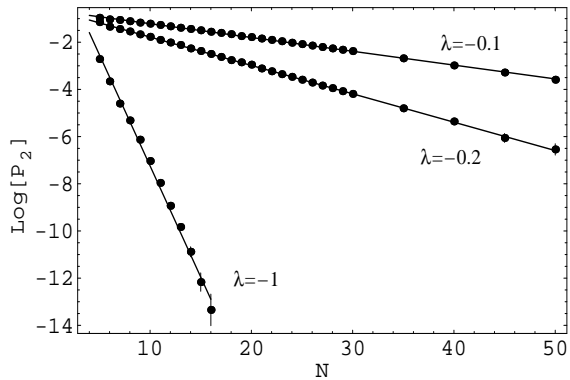} 
\end{center}
\caption{\small  Numerical
  values of the logarithm of the probability that a $2$-cycle
 appears in the optimal tour as a function of $N$
 for  $\lambda=-1, -0.2$ and $-0.1$.}
\label{fig3}
%\end{ruledtabular}
\end{figure}
%%%%%%%%%%%%%%%%%%%%%%%%%% 
%% \ldots.
%% (COMENTAR LA FIGURA)
As can be seen in Fig. \ref{fig3}, the
 probability of having a two-cycle
 in the optimal solution decays exponentially
 with $N$ with a coefficient 
 that depends on $\lambda$. For $\lambda =-1$  the points can be fitted to 
 $\log [P_2] =2.17653 - 0.941985 N$, i. e. $P_2\sim A\xi^{-N}$
with $\xi=2.565...$, so when
 $N > 20$ we can neglect the $2$-cycles and then all the distances that
 appears in the optimal tour will be
 uncorrelated. The problem is again reduced
 to the study of the subset of the permutation group, in this case  
without  1-cycles and 2-cycles. 
This can be done
 with the help of the associated Stirling numbers $d_3(N,k)$ and their 
 generating function. Finally for $\lambda=-1$, the value of $\langle n_{c}\rangle$ can be
 computed (up to corrections of order $\xi^{-N}$) as:
\begin{eqnarray}
\langle n_{c}\rangle_{_{\lambda=-1}} & = & -{ \left[ {d^N\over dx^N} \left( {(\log (1-x)+p(x) )\exp (-p(x))
\over 1-x}\right)\right]_{x=0} \over  \left[ {d^N\over dx^N} \left( { \exp (-p(x))
\over 1-x}\right)\right]_{x=0}}\nonumber \\
 & = & H_N-{3\over2}+ \sum_{i=1}^{\infty} C_i N^{-i},
\label{d3}
\end{eqnarray}
where $p(x)=x+x^2/2$, and the first coefficients
 in (\ref{d3}) are $C_i= 2,$ $-3/2,$ $ -5/6,$ $
7/4, $ $106/15,$ $67/12,$ $-2627/42,$ $-8633/24,$
 $47929/90,31758/5,$ $1989059/33 $ for
$i=1,\dots,11$.  In the
 inset of Fig. \ref{fig1} we plot, together,
 the values of $\langle n_{c}\rangle/\log(N)$  obtained form (\ref{d3}) 
 and the experimental points, so we can see that the agreement 
 is excellent. It should be noted that in the limit $N \to \infty$
 we obtain from (\ref{d2}) and (\ref{d3}) the relation 
\begin{equation}
\lim_{N\to \infty} (\langle n_{c}\rangle_{_{\lambda=0}} - \langle n_{c}\rangle_{_{\lambda=-1}} ) ={1\over2}.
\label{d4}
\end{equation}
which states that, for large $N$, between $\lambda =0$ and $\lambda=-1$
 the value of $\langle n_{c} \rangle $ decreases only  $0.5$ units. 
Actually we can show that
\begin{equation}
\lim_{N\to \infty} (H_N-\langle n_{c}\rangle_{_\lambda})={3\over 2}\quad  
\forall\lambda\in[-1,0)
\end{equation}
which tells us that in the thermodynamic limit $N\to\infty$
the expected value of the number of cycles is independent of $\lambda$ in the 
range $\lambda<0$.
The theoretical explanation for this fact is again based in the 
exponential suppression of  $2$-cycles shown in Fig. 3, 
and therefore the equal probability
of all permutations (without $1$ or $2$-cycles) for producing the
optimal tour. This, in turn, implies that up to corrections
of order ${\cal O}(\xi^{-N})$ the results are independent of 
the probability distribution as it happened in the random ($\lambda=0$) 
case. In this sense the concrete distribution used for the entries of 
our random matrix (\ref{ranmat}) 
that except for $\lambda=0$ is non uniform,  
has little effect on the results 
and the influence weakens more and more 
as the dimension $N$ grows.

 %%%%%%%%%%%%%%%%%%%%%%%%%%%%%%%%%%%%%%%%%%%%%%%%%
\begin{figure}[h!]
%\begin{ruledtabular} \hsize=8.5cm
\begin{center}
\includegraphics[width=7.5cm,height=5.cm]{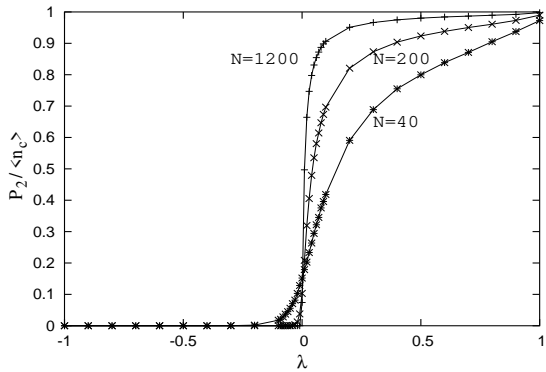} 
\end{center}
\caption{\small  Plot of the ratio between the number 
 of 2-cycles ($P_2$) and the total number of cycles versus $\lambda$ for 
different values of the dimension.}  
\label{fig4}
%\end{ruledtabular}
\end{figure}
%%%%%%%%%%%%%%%%%%%%%%%%%% 

The right half of figure 1, corresponding to positive values of $\lambda$,
is much more poorly understood.
It is clear from the diagram that $\langle n_{c}\rangle$ behaves linearly 
with $\lambda$ with a slope close to $N/2$. This linear behavior has corrections
near $\lambda=1$ that are visible, for instance, in the change of the slope
 in the curve for
 $N=200$ at $\lambda=0.8$. It is apparent that
the cycles of the optimal solutions in this side are dominated by 2-cycles,
and as one can see from Fig. 4
$$\lim_{N\to\infty}{P_2\over\langle n_{c}\rangle_{_{\lambda}}}=1\quad \forall\lambda>0.$$
Actually the number of $2$-cycles grows linearly with $N$ for all
positive values of $\lambda$ while the rest, i. e. the  total number 
of $k$-cycles with $k\not= 2$ grows logarithmically. As 
$\lambda$ approaches $0$ from the positive side the linear
behavior with $N$ in $P_2$ has a smaller slope and 
the average number of $2$-cycles goes to a constant $1/2$ 
(for large $N$)  in the $\lambda=0$ limit.  
 
In relation to this we can prove that in the symmetric point
($\lambda=1$) the probability of having a $2k$-cycle 
with $k>1$ vanishes. For if the links in
even positions of the cycle
add up to a length smaller (greater) that those in odd positions
then breaking the $2k$-cycle into $k$ $2$-cycles,
using the even (odd) links, lowers the length of the tour.
As a result instead of a $2k$-cycle we would have $k$ $2$-cycles
in the optimal tour. This feature helps to understand  
the abundance of $2$-cycles for positive values of $\lambda$ and the change of the
slope near $\lambda = 1$.

 %%%%%%%%%%%%%%%%%%%%%%%%%%%%%%%%%%%%%%%%%%%%%%%%%
\begin{figure}[h!]
%\begin{ruledtabular} \hsize=8.5cm
\begin{center}
\includegraphics[width=7.5cm,height=5.cm]{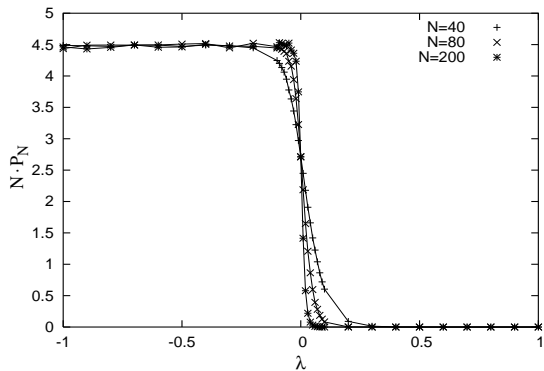} 
\end{center}
\caption{\small  Plot for the probability ($P_N$)
of having an optimal solution consisting of one $N$-cycle
times the dimension $N$ versus $\lambda$.}
\label{fig5}
%\end{ruledtabular}
\end{figure}
%%%%%%%%%%%%%%%%%%%%%%%%%% 

The {\it dimerized} phase corresponding to $\lambda>0$ contrasts
with the {\it polymerized} one for $\lambda<0$. 
% As shown in Fig 5, 
In this region there are $N$-cycles in the optimal solution with a 
probability $P_N$ obtained from the corresponding generating function
\begin{equation}
\label{PN1}
P_N(\lambda<0)={{\rm e}^{3/2}\over N}+{\cal O}(\xi^{-N})
\end{equation}
 while $N$-cycles are exponentially suppressed in 
the dimerized region. The intermediate case corresponds
to $\lambda=0$ with
\begin{equation}
\label{PN2}
P_N(\lambda=0)={{\rm e}\over N}+{\cal O}(1/N!).
\end{equation}
The error in (\ref{PN1}) is produced by the residual presence of 
2-cycles as discussed after Fig. 3, 
while that of (\ref{PN2}), much smaller,  
comes from the tail of the series for the exponential.
This has been verified numerically, the results are
shown in Fig. 5. A more detailed analysis of
 the scaling as well as the
``critical'' properties of the transition is in progress.

Notice that in the polymerized phase ($\lambda<0$) 
there are a non-negligible 
probability of finding a solution of the TSP in polynomial time,
actually after $N$ runnings of the problem one would solve
in a time of the order $N \cdot N^3$ a number of TSP's approximately equal to
${\rm e}^{3/2}=4.48...$ 

In conclusion, we have found
 a phase transition in the random assignment problem that
 separates two regimes where,
 while remaining always solvable in polynomial time,
 the system approaches the
 traveling salesman problem and the simple matching problem
 respectively. This phase transition can
 be seen as complementary of that 
 which appears in many NP-complete problems separating
 the easy instances from the hard ones 
 (\cite{Monasson}, \cite{Mertens}). In our case, 
 the control parameter is the correlation 
 between the distances $d_{i,j}$ and $d_{j,i}$ which in turn
 controls the existence,
 or not, of allowed 2-cycles in the optimal tour.
 When both distances are positively correlated, the $2$-cycles
 are allowed and favored and in the limit $N\to \infty$, 
 the $\langle n_c\rangle$
 is dominated by the $2$-cycles as seen in Fig. \ref{fig4}, consequently,
 the AP optimal solution for those instances
 is very far away of that of the TSP solution
 for the same matrix. On the contrary,
 when $d_{i,j}$ and $d_{j,i}$ are negatively correlated it
 is very unlikely that a $2$-cycle enters in the optimal solution,
 consequently they are strongly suppressed
 (see Fig.-\ref{fig3} and \ref{fig4}) and the optimal tour
 is composed of a ``few''  cycles ($\ \approx \log(N)$) which means that the AP
 optimal tour for those distances is near the one of the TSP
 problem for the same matrix. This property can be useful as a starting point 
 for designing improved algorithms for solving the traveling salesman 
 problem. 
%%%%%%%%%%%%%%%%%%%%%%%%%%%%%%%%%%%%%%%

\acknowledgments{We thank Julio Abad  
for useful discussions.
This work has been supported the
CICYT  grants
  BFM2003-01300 and FPA2003-02948.
}
%%%%%%%%%%%%%%%%%%%%%%%%%%%%%%%%%%%%%%%%%%%%%%%%%%%%%%%%%%%%%
%%%%%%%%%%%%%%%%%%%%%%%%%%%%%%%%%%%%%%%%%%%%%%%%%%%%%%%%%%%%%%%%%%


\begin{thebibliography}{99}
\bibitem{dist}{Note that $d_{i,j}$ are not true distances, in a mathematical
    sense,
 in particular  
they do not need to be positive or symmetric.}
\bibitem{neg}{Note that for negative values of $\lambda$, some ``distances''
 can be negative. This could be avoided by adding a constant $\lambda$ term
 to all matrix elements $d_{i,j}$, but this only changes the length of the
 minimal  path in a constant factor $N \lambda$ and does not change the
 number of cycles.}
\bibitem{prueba}
{To see this, consider two allowed permutations
$\sigma$ and $\sigma '$ and the map $T: d \to \tilde{d}$ such that  $\tilde
{d}_{i,\sigma'(i)}=d_{i,\sigma(i)} $, $\tilde {d}_{i,\sigma(i)}=d_{i,\sigma'(i)} $ and
$\tilde {d}_{i,k}=d_{i,k} $, ($k\neq \sigma (i), \sigma'(i); i=1,N $). For $\lambda =0$, T leaves
 invariant the probability distribution for $d$ and transforms
 $\sigma$ into $\sigma '$. Then both permutations have equal probability of leading to
 the shortest path.}
\end{thebibliography}
\end{document}